\documentclass[a4paper,11pt]{article}
\pdfoutput=1 

\usepackage{jcappub} 

\usepackage[T1]{fontenc} 
\usepackage{graphicx}
\usepackage{txfonts}
\usepackage{xcolor}
\usepackage{amsmath} 
\usepackage{hyperref}
\usepackage{aas_macros}
\usepackage{threeparttable}

\title{3C\,403: a candidate neutrino-emitting radio galaxy}

\author[a]{Gabriele Bruni}
\author[b]{Loredana Bassani}
\author[c]{Sergio Alves Garre}
\author[d]{Manuela Molina}
\author[b]{Angela Malizia}
\author[a]{Mariateresa Fiocchi}
\author[a]{James Rodi}
\author[e]{Antoine Kouchner}
\author[e]{Alexis Coleiro}
\author[e]{Julien Aublin}
\author[f]{Giulia Illuminati}
\author[a]{Francesca Panessa}
\author[a]{Angela Bazzano}
\author[a]{Lorenzo Natalucci}
\author[a]{Pietro Ubertini}

\affiliation[a]{INAF -- Istituto di Astrofisica e Planetologia Spaziali,\\ via del Fosso del Cavaliere 100, Roma, 00133, Italy}
\affiliation[b]{INAF -- Osservatorio di Astroﬁsica e Scienza dello Spazio,\\ Via Piero Gobetti 93/3, I-40129 Bologna, Italy}
\affiliation[c]{IFIC -- Instituto de Física Corpuscular (CSIC - UV),\\
C/ Catedrático José Beltrán nº2,
E-46980, Paterna, Spain}
\affiliation[d]{INAF -- Istituto di Astrofisica Spaziale e Fisica cosmica,\\ via Corti 12, I-20133 Milano, Italy}
\affiliation{INAF -- Istituto di Astrofisica e Planetologia Spaziali,\\ via del Fosso del Cavaliere 100, Roma, 00133, Italy}
\affiliation{INAF -- Istituto di Astrofisica e Planetologia Spaziali,\\ via del Fosso del Cavaliere 100, Roma, 00133, Italy}
\affiliation[e]{Universit\'e Paris Cit\'e, CNRS, Astroparticule et Cosmologie,\\ F-75013 Paris, France}
\affiliation[f]{INFN, Sezione di Bologna,\\ v.le C. Berti-Pichat, 6/2, Bologna, 40127 Italy}

\emailAdd{gabriele.bruni@inaf.it, loredana.bassani@inaf.it, salves.phys@gmail.com, manuela.molina@inaf.it, angela.malizia@inaf.it, mariateresa.fiocchi@inaf.it, mariateresa.fiocchi@inaf.it, kouchner@apc.in2p3.fr, coleiro@apc.in2p3.fr, julien.aublin@apc.in2p3.fr, giulia.illuminati3@unibo.it, francesca.panessa@inaf.it, angela.bazzano@inaf.it, lorenzo.natalucci@inaf.it, pietro.ubertini@inaf.it}

\abstract{
3C\,403 is a well-known FR\,II radio galaxy with jets extending up to kiloparsec scales. 
We report its identification as the second most significant candidate among more than 150 sources examined using the 15-year neutrino dataset from the ANTARES Collaboration, making it one of the most promising radio-galaxy candidates for high-energy neutrino emission. 
Motivated by previous associations between blazars and neutrino events, we investigated the jet properties of 3C\,403 and their possible role in neutrino production. 

Multi-scale radio observations, from parsec to kiloparsec scales, reveal a stable, two-sided jet lying close to the plane of the sky, with no evidence of strong Doppler boosting, while X-ray data indicate a dominant, heavily absorbed accretion-related component. 
We also examined the recently proposed correlation between neutrino and hard X-ray fluxes—originally identified in blazars and Seyfert galaxies—and find that 3C\,403 occupies an intermediate location in the $L_{\nu}$--$L_{\rm hX}$ plane between jet-dominated and corona-dominated systems. 
However, the current upper limit on its neutrino flux prevents a firm assessment of whether it follows the proposed relation.

With radiatively efficient accretion ($\lambda_{\rm Edd}\sim10^{-2}$), strong hard X-ray emission, and a powerful but misaligned jet, 3C\,403 provides a physically motivated laboratory for exploring the interplay between coronal activity and jet environments in multimessenger scenarios of neutrino production in active galaxies.

}

\begin{document}
\maketitle
\flushbottom

\section{Introduction}
In 2013, the IceCube collaboration announced the detection of an excess of neutrinos above the expected background level showing  a spatial distribution compatible with isotropy, and thus with an extra-galactic origin \citep{PhysRevLett.111.021103}. A few years later, the same collaboration  announced the detection of  an extremely-high-energy neutrino  traced back to the blazar TXS\,0506+056 \citep{doi:10.1126/science.aat2890}, during a period of intense gamma-ray emission observed by \textit{Fermi}/LAT and MAGIC telescopes in September 2017 \citep{2019NatAs...3...88G}. Moreover, analysing  previous IceCube data  obtained  from  October 2014 to February 2015, a 3.5$\sigma$ excess  neutrino emission was also found from the same source direction  \citep{doi:10.1126/science.aat2890}. However, this last neutrino excess was not accompanied by gamma-ray flaring activity \citep{2019ApJ...880..103G}, implying  a complex relation  between  GeV photons/Tev neutrinos fluxes and/or  some peculiarity  of the source properties.

Two models were invoked for neutrino emission: a photohadronic ($p\gamma$) one, foreseeing proton interaction with ambient photons (from the accretion disk, synchrotron photons emitted from the jet, CMB photons strayed into the jet) and a hadronic ($pp$) one, where protons interact with other protons within the corona, the jet, or with protons of the external material trapped in the jet flow \citep{1989A&A...221..211M,1990ApJ...362...38B,PhysRevLett.116.071101,PhysRevLett.125.011101} 
However, without an obscuring mechanism at work in most or all such sources, the $\gamma$-rays accompanying these processes would outshine the cosmic gamma-ray background measured by the \textit{Fermi}/LAT \citep{PhysRevLett.132.101002,PhysRevLett.125.011101}.

The second source with a neutrino association was the Seyfert galaxy NGC\,1068 \cite{doi:10.1126/science.aat2890}. The discrepancy noted between neutrinos and $\gamma$-rays fluxes from NGC\,1068 could not be accounted for by absorption via the extragalactic background light (EBL) \citep{Murase_2022}. Consequently, the regions within NGC 1068 where the neutrinos originate must be able to obscure the GeV–TeV $\gamma$-rays expected to be emitted along with the neutrinos. This is reflected by the well known presence of a highly-obscured, Compton-thick nucleus in this source \citep{1997A&A...325L..13M,2015ApJ...812..116B}. The surroundings of the central black hole, i.e. the corona, suits well in this scheme, offering a hadron-rich environment where both neutrino production and $\gamma$-rays obscuration can take place. Indeed, models were proposed in the previous years where the corona was seen as the production site for neutrinos 
\citep{PhysRevLett.116.071101,Murase_2022,2020ApJ...891L..33I,2021ApJ...922...45K, 2024ApJ...972...44D,2024PhRvD.109j1306M,2025arXiv250406336F,2024ApJ...961L..14F,2024ApJ...974...75F,2025arXiv250201738F,2025arXiv250406336F}, 
and foreseeing a linear correlation between the unabsorbed hard X-rays and neutrino fluxes \citep{PhysRevD.94.103006}, where NGC\,1068 was already indicated as one of the best candidates for neutrino association \citep{PhysRevLett.125.011101}. These were recently invoked to reinterpret the neutrino production in blazar TXS\,0506+056 without the need for a jet contribution \citep{2024PhRvD.110l3014K,2025arXiv250201738F}.

Recently, further sources have been associated with neutrino events detected by the IceCube observatory: two
blazars (PKS\,1502+106 and PKS\,1424-41, \cite{2019ATel12967....1T,2016NatPh..12..807K})
and three Seyfert galaxies (NGC\,4151, NGC\,3079, and
CGCG\,420-015, \cite{PhysRevLett.132.101002,2024arXiv240607601A}). This increased statistics allowed for the first time to test the predicted hard X-rays / neutrino flux correlation: \cite{2024PhRvD.110l3014K} showed how the two blazars GB6\,J1542+6129 and TXS\,0506+0566, despite their powerful jets, lie on the expected correlation together with Seyfert galaxies NGC\,1068, NGC\,3079, NGC\,4151, and CGCG\,420-015, pointing to a possible common origin for the neutrinos. 
Interestingly, \cite{Ambrosone2024} revisited the Berezinsky scenario of hidden neutrino sources, deriving new constraints on the contribution of Seyfert galaxies to the diffuse neutrino background. Their analysis shows that, under sub-equipartition between cosmic-ray and magnetic energy densities in hot coronae, Seyfert galaxies can account for part of the observed low-energy neutrino flux, while remaining consistent with the non-detection above $\sim$100,TeV.

A further way to test the role of the jet in neutrino production is to study the possible association with jets lying on the plane of the sky, thus not presenting a privileged point of view on the ongoing processes in inner regions of the jet (thus non-blazar). This approach may lead to a refinement of the actual models, allowing to test the role of alternative production regions. 
In this work, we present a radio to hard X-rays study of the first candidate radio galaxy (Seyfert 2) possibly associated with neutrino events from the ANTARES experiment. In the following, we discuss the background from previous studies on this source (Sec. \ref{3c403}), the collected multiwavelength data (appendix \ref{data}), the jet and corona properties (Sec. \ref{jet} and \ref{corona}), and the possible agreement with the hard X-rays versus neutrino flux correlation (Sec. \ref{kun}).  

We adopt a $\Lambda$CDM cosmology with $H_0=67.7$ km/s/Mpc, $\Omega_\Lambda=0.691$, and $\Omega_M=0.307$. At the redshift of the source ($z=0.058$), the physical scale is 1.168 kpc/".


\section{The radio galaxy 3C\,403}
\label{3c403} 
Recently, the  ANTARES collaboration used data collected between 2007 and 2018 to find possible correlations with several AGN catalogues, including a sample of soft $\gamma$-ray selected radio galaxies compiled by \cite{2016MNRAS.461.3165B}.  They  found the presence of 2 events (of $\sim$5 TeV and $\sim$10 TeV, respectively)  located at less than 0.5 degrees from the radio galaxy 3C\,403,  implying a pre-trial p-value equivalent to a 3.7$\sigma$ excess, and to 2.5$\sigma$ post-trial \citep{2021ApJ...911...48A}. In the final 15-year candidate-list analysis performed by the collaboration this source remains significant as the second best candidate for a neutrino association, with three events closer than $1^\circ$ and a pre-trial p-value of 3.4$\sigma$~\citep{AlvesPhD_2025}.

\begin{figure*}
    \centering
    \begin{minipage}{1\linewidth}
        \centering
        \includegraphics[width=\linewidth]{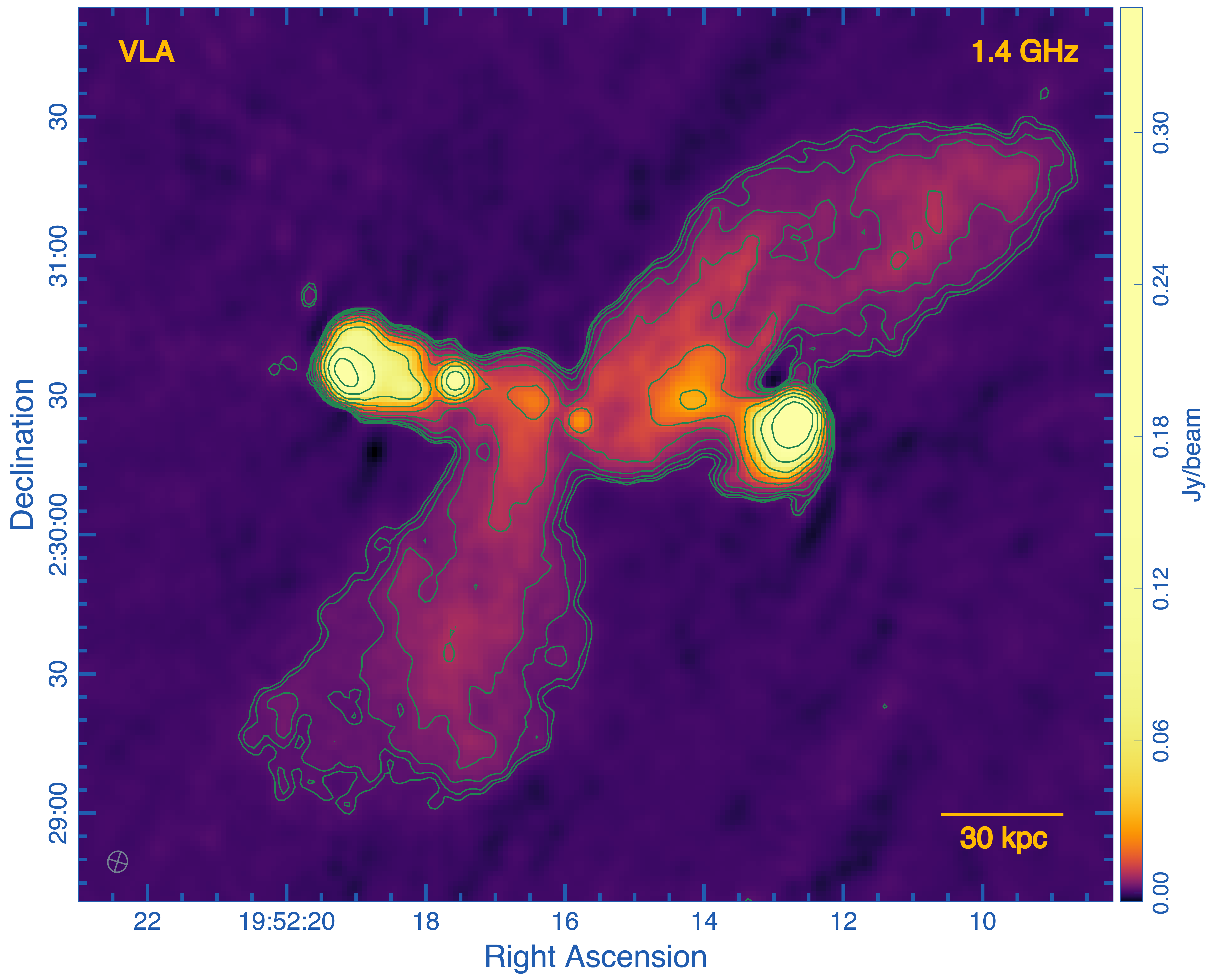}
    \end{minipage}

    \vspace{0.5cm} 

    \begin{minipage}{1\linewidth}
        \centering
        \includegraphics[width=\linewidth]{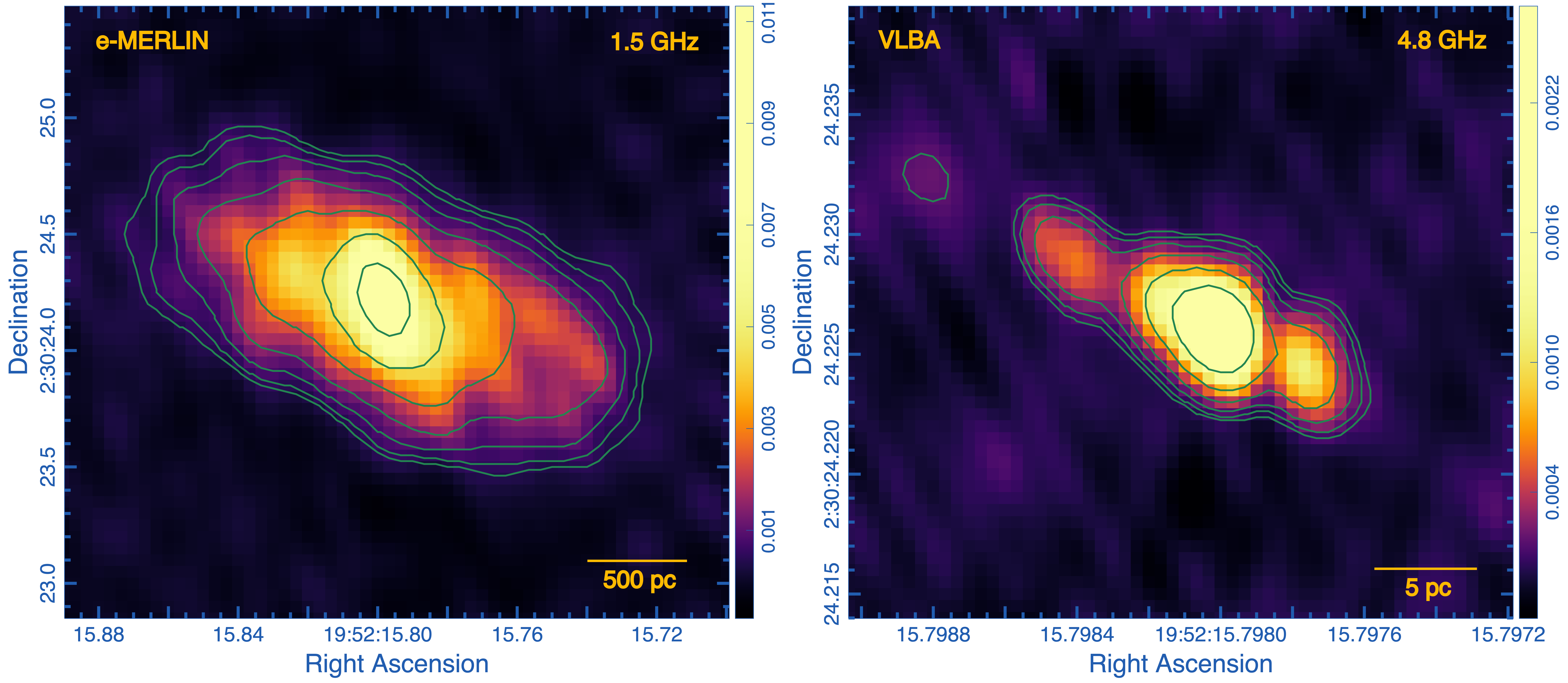}
    \end{minipage}
    \caption{Multi-scale radio images of 3C\,403, from kiloparsec to parsec scales. Top: archival VLA image (NVAS) at 1.4\,GHz showing the characteristic winged morphology of the source. Bottom left: e-MERLIN image at 1.5\,GHz resolving the inner kiloparsec of the jet. Bottom right: VLBA image at 4.8\,GHz tracing the central $\sim$20\,pc. The jet position angle is consistent across all spatial scales. Contours increase exponentially as $3\times{\rm RMS}\times2^{n}$.}
\label{fig:radio}
\end{figure*}

This galaxy, at redshift 0.0589, is from many points of view peculiar: it is a type 2 radio galaxy with high excitation emission lines which hosts megamaser emission (the only one detected in an FR\,II radio galaxy so far - see \cite{2003A&A...407L..33T,2007A&A...475..497T,2020A&A...641A.162P}). The source is listed as a persistent object, in both \emph{Swift}/BAT \citep{2023AAS...24125407L} and \textit{INTEGRAL}/IBIS \citep{Bird_2016} catalogues, which collect data from the past $\sim$20 years, while it remains undetected by \textit{FERMI}/LAT in the 4FGL-DR3 catalogue \cite{Abdollahi_2022}. In the radio band, it is classified as an FRII type galaxy with a peculiar morphology, called winged or X-shaped (see Figure~\ref{fig:radio}) in which two pairs of lobe-like features are present: wings are less luminous, diffuse, and have a steep spectrum emanated from slowly expanding plasma, while the main lobes are brighter and connected to the ongoing jet, which replenishes them with energetic particles (see e.g. \cite{2007AJ....133.2097C,2018ApJ...852...48S}). 

Possible mechanisms for the formation of the X-shaped or winged  morphology can be: 1) a  change in the jet axis due to the activation of a new radio phase with a different orientation - possible signature of a binary black hole merger or the merger with a smaller galaxy \citep{1980Natur.287..307B}; 2) backflow of plasma from the lobes towards the core, deflected from the host galaxy halo (see e.g. \cite{2020MNRAS.495.1271C}; 3) a relatively slow conical precession of the jet axis, resulting in the observed X-shaped morphology by projection \citep{1985A&AS...59..511P,1994A&AS..103..157M}.
In the case of 3C 403, \cite{2002MNRAS.330..609D} discussed the evidence against a merger: the source is located in a sparse environment, and an analysis of optical continuum and emission-line images from the literature showed no sign of disturbances. The same authors disfavour a conical precession as well: a structure with relatively high surface brightness connecting the wing and lobe would likely appear if the morphology resulted from a specific alignment of a slowly precessing source. Additionally, the constriction of the wings at their base also suggests that interpretations involving slow jet axis motion are unlikely. Finally, the lack of spectral curvature up to 32 MHz in the wings poses problems that the plasma age is too young to be due to a previous radio phase. Alternatively a very fast jet reorientation would be need (a few Myr, \cite{2002MNRAS.330..609D}), while a backflow directly connected to the jet would more naturally explain the lack of steepening. 

\cite{2005ApJ...622..149K} performed imaging analysis of the first \textit{Chandra} observations of this X-shaped radio galaxy, and found that the X-ray emission from 3C\,403 host galaxy is highly elliptical and co-aligned with the optical isophotes. This supports the hypothesis that X-shaped radio sources emerge from the propagation of jets through uneven density regions. Furthermore, the same authors found that there is no indication that the lobes or wings are more pressurized compared to the interstellar medium (ISM), which supports the hypothesis that the wings arise from robust backflows of material from the jet head followed by buoyant evolution. The latter scenario is also supported by more recent studies on similar sources, making use of new generation low-frequency radio telescopes such as LOw-Frequency ARray (LOFAR) able to recover the extended regions generated by buoyancy effects. Indeed, \cite{10.1093/mnras/stz1910} showed how the wings of the prototypical X-shaped source NGC\,326 result far more extended and complex than the actual lobes when observed in the MHz domain, eventually suggesting a hydrodynamical origin rather than a jet precession. Unfortunately, due to its equatorial declination 3C\,403 does not fall in the footprint of the LOFAR Two-metre Sky Survey (LoTSS), not allowing us to perform a similar study.  

Given these premises, and to further characterise the jet properties of this peculiar source, we collected new multi-scale radio data from the kpc to pc scale, allowing us to assess the jet orientation and evolution along a Myr time scale.


\section{Jet properties from kpc to pc scale}
\label{jet}

We explored the pc to kpc scale radio properties of the source through multi-scale, multi-resolution observations. Derived quantities are reported in table \ref{tab:radio}. The central region, including the core and the inner kpc of the jet and counterjet, is clearly visible in our e-Merlin 1.5 GHz image (see Fig. \ref{fig:radio}, bottom-left panel). The position angle of the jet ($70^\circ\pm5^\circ$) is consistent with that of the axis connecting the lobes visible in the larger scale VLA image ($72^\circ\pm2^\circ$, see top panel of Fig. \ref{fig:radio}). The radio emission detected by e-Merlin (jets plus core) has a total angular size of 2.4 arcsec, corresponding to 2.8 kpc. 

In addition, we carried out VLBA observations to zoom into the central few parsecs of the jet: also on this scale, both jet and counterjet are clearly visible in the 5 GHz (see Fig. \ref{fig:radio}, bottom-right panel) and 8 GHz images (see Appendix), excluding a line of sight close to the jet axis. At 15 GHz, only the core is detected (see Appendix). The jet position angle as measured at 5 GHz ($64\pm5$ deg) is consistent with ones measured with e-Merlin and VLA, excluding a jet precession on the overall pc to kpc scale.
A previous VLBI study of the source was presented by \cite{2007A&A...475..497T}, using EVN observations at the same frequency (4.8 GHz) and performed in 2004. Despite the lower angular resolution along the jet, those authors could estimate a position angle consistent within errors with the multi-scale observations presented above ($72.5\pm3$ deg). 

Our VLBA and e-Merlin observations (together with archival VLA ones) recovering the jet emission and morphology from the kpc to pc scale, confirm the absence of jet beaming and the agreement of jet axis on the different scales, not suggesting the presence of precession. Indeed, the symmetric jet/counterjet morphology  -- even on the pc scale -- excludes beaming, while the consistent position angle of the jet on different scales rules out precession on a $\sim$Myr timescale. We can attempt to quantify the lack of beaming using the jet--counterjet ($R_{\rm j/cj}$) brightness ratio from our VLBA observation, although some assumptions on the jet kinematic are needed. For such a purpose, we consider the well known relation \citep{1998ApJ...493..632G}: 
\begin{equation}
R_{\rm j/cj} \;=\; 
\left(\frac{1+\beta\cos\theta}{1-\beta\cos\theta}\right)^{2+\alpha},
\label{eq:Rj}
\end{equation}
with $S_\nu\!\propto\!\nu^{-\alpha}$. 
From our VLBI images we measure $R_{\rm j/cj}\sim1.3$ at 5\,GHz. 
We adopt $\alpha=0.6$, appropriate for optically thin synchrotron emission in inner FR\,II jets at cm wavelengths (i.e. away from the flat-spectrum core). 
For representative mildly relativistic speeds ($\beta=0.6$--0.8) we obtain 
$\theta \simeq 85^\circ$.
These values indicate a jet very close to the plane of the sky and thus negligible Doppler boosting, 
consistent with the observed two-sided morphology and the low core dominance.

Overall, these results hint to a different neutrino-producing mechanism with respect to the ones involving the jet initially proposed for the blazar TXS\,0506+056. In the next sections, we explore the high-energy properties of 3C\,403, characterizing its corona emission, and its possible link with neutrino emission.


\section{High energy properties}
\label{corona}
3C 403 has been extensively observed at high energies from soft to hard X-rays over the course of several years. 
As already mentioned, 3C 403 was observed by Chandra
in 2002 and thoroughly analysed by \cite{2005ApJ...622..149K}, who analysed the X-ray emission coming from
both the active nucleus and from several radio components, including lobes and wings and compact regions in the jet.
The nuclear component is well described by two power-laws plus a fluorescent iron line from cold material. The first power-law has a photon index of 1.7 and is heavily absorbed ($\rm{N_H}\sim4\times10^{23}\,\rm{cm}^{-2}$) as expected from coronal emission in a type 2 AGN. The second power-law  is less absorbed and  its luminosity is more than 100 times weaker than the primary continuum. This extra component has been interpreted by \cite{2005ApJ...622..149K} as X-ray emission from an unresolved jet.
This is supported by the observed value of the X-ray-to-radio flux ratio which is consistent with what generally found in low-power radio galaxies, 
by the much lower level of absorption in the jet with respect to the torus and also by the presence of a pc-scale jet near the core at radio wavelenghts. 
Both the corona and the unresolved jet  could be the sites of neutrino emission, but the dominance of the first with respect to the other combined with  the conclusions reached in the previous section  suggests  the corona as the most likely site. The dominance of the absorbed coronal X-ray component and the lack of significant flux or absorption variability suggest that the bulk of the high-energy emission arises from a compact, quasi-steady region close to the black hole. 
Such conditions are consistent with a hot, magnetized corona where protons can be accelerated to relativistic energies and efficiently interact with ambient X-ray photons via $p\gamma$ processes 
\citep[e.g.][]{PhysRevD.94.103006,Inoue2020,Fiorillo2024b}. 
Assuming coronal densities of $n_{\mathrm{H}}\!\sim\!10^{9-10}\,\mathrm{cm^{-3}}$ and magnetic fields of $B\!\sim\!10^{2-3}\,$G, as inferred for luminous Seyfert nuclei, the mean free path for $p\gamma$ interactions becomes comparable to the coronal scale height, allowing an efficient conversion of cosmic-ray energy into neutrinos. 
In contrast, the misaligned radio jet of 3C\,403, with negligible Doppler boosting and limited radiative output at high energies, appears less favorable for neutrino production through beamed hadronic or photohadronic processes typically invoked in blazars. 
We therefore identify the magnetized corona as the most likely site of neutrino emission in this source, consistent with recent models developed for NGC\,1068 and TXS\,0506+056.

To characterise 3C\,403 nuclear emission in more detail, we analysed a 2013 contemporaneous Swift/XRT and NuSTAR observation, fitting the data with a double power-law 
plus an iron line as done by \cite{2005ApJ...622..149K} (see Appendix for details on data  reduction). X-ray spectral analysis was performed using XSPEC v.12.13.1 (Arnaud et al. 1996) and spectra were binned with a minimum of 20 counts per bin in order to use the $\chi^2$ statistics; all errors are quoted at 90\% confidence level ($\Delta\chi^2$=2.71 for one parameter of interest). The model used, in \texttt{XSPEC}
terminology is \texttt{const*phabs[po+phabs*(po+zga)]}, where the first absorption component is the
Galactic column density, while the second refers to absorption intrinsic to the source, the first
power law is used to approximate the jet contribution, whereas the second power law describes the 
primary continuum emission. The iron line, described by a Gaussian component, is compatible with fluorescent emission from cold material and its EW ($\sim$200 eV) is consistent with a model in which the line emission originates from an absorbing region
that surrounds the central supermassive black hole \cite{10.1093/pasj/48.6.801}. As can be seen from table \ref{best_fit}, our results are compatible with what they found in the Chandra study. 
Assuming that the two power-law 
components represent the corona and the jet emission in 3C\,403, we are able to measure both at high energies for the first time and confirm that the primary 
power-law overwhelms the latter by a large factor even higher energies ($E>20$ keV). As a further step, we tested the broad-band spectrum against the presence of a high energy cut-off and a reflection component, but neither of them are properly constrained, resulting in a lower limit of the cut-off energy at around 50 keV and an upper limit for the reflection fraction of $\sim$1.3.  These findings, together with the flat photon index and the relatively small EW of the iron line, are in line with the X-ray properties of radio galaxies  which, on average, show less prominent reflection features than normal AGN (see for instance \cite{2007MNRAS.382..937M,2013MNRAS.428.2901W,2020ApJ...901..111K}). The measured EW is fully consistent with the upper limit on the reflection fraction, as discussed by \cite{2019MNRAS.484.2735M}.
Moreover, the reflection parameter is inherently difficult to constrain given the limited quality of the available data.
The fact that only an upper limit is obtained does not rule out the presence of reflection; rather, it suggests that its strength may be lower than unity, in agreement with the EW value derived.
  
While \textit{Swift}/XRT and NuSTAR data represent a snapshot of the source state in 2013, the \textit{INTEGRAL}/IBIS and \textit{Swift}/BAT ones can provide average information over a period of almost two decades. For IBIS, we summed observations performed over the period between March 2003 and September 2024: the 15-55 keV spectrum, which is well described by a simple power-law with $\Gamma=2.4\pm1.0$, results in an unabsorbed flux of $1.8\pm0.4\times10^{-11}\ \mathrm{erg\ cm^{-2}\ sec^{-1}}$). The BAT spectrum\footnote{https://swift.gsfc.nasa.gov/results/bs105mon/}, covering a slightly different period (Dec. 2004 - Dec. 2017), provides very similar results ($\Gamma=1.8\pm0.2$ and a 15-55 keV flux of $1.2\pm0.2\times10^{-11}\ \mathrm{erg\ cm^{-2}\ s^{-1}}$ ). Both are consistent with the single epoch, \emph{NuSTAR} one in the same band ($1.58\pm0.47 \times10^{-11}\ \mathrm{erg\ cm^2\ s^{-1}}$), excluding strong variability in the hard X-rays band.

To check if variability is instead present at softer energies, we have collected soft X-ray data (2-10 keV) from the literature (see table \ref{x_ray_fluxes}), including 
Suzaku data taken in 2009 in combination with \textit{Swift}/BAT \citep{2011ApJ...738...70T} and XMMslew  data taken in 2010\footnote{https://www.cosmos.esa.int/web/xmm-newton/xsa}. 
3C\,403 has also been observed by \textit{Swift}/XRT several times in the period between 2006 and 2013, and then lastly in June 2023 and at the end of 2024.  
All data have been reduced following the prescription reported in the appendix and analysed as described above.

In Table \ref{x_ray_fluxes} we report the flux and absorption history of 3C\,403, by listing all the available measurements. As can be seen from the value reported in the table, the source is not characterised by evident variability either in flux or absorption properties, taking into account the large uncertainties on both parameters. 
All 2-10 keV fluxes are compatible with an average value of  3$^{+1.1}_{-0.7}$$\times$10$^{-12}$ erg\,cm$^{-2}$\,s$^{-1}$, i.e similar to what measured by NuSTAR. Also the $\rm{N_H}$ values are consistent with an average value of 3.4$^{+1.3}_{-0.9}\times10^{23}$ atoms cm$^{-2}$, again similar to the NuSTAR result. This suggests that 3C\,403 is
not highly variable at X-ray energies (with at most a 30-40\% variations on both parameters), and that the XRT and NuSTAR observations are representative of the source flux state.


\begin{table}
 \caption{Soft X-ray flux and absorption history of 3C\,403.}             
 \label{xray_fluxes}     
 \centering     
 \begin{threeparttable}
 \begin{tabular}{l c c c }     
 \hline\hline                 
 Date & Telescope & F$_{\rm 2-10 keV}$& N$_{\rm H}$\\
    &             &10$^{-12}$erg cm$^{-2}$ s$^{-1}$& 10$^{23}$cm$^{-2}$\\
 \hline                        
Dec. 2002       & Chandra & 1.1\tnote{$\dagger$}&4$\pm$2\\
Oct./Nov. 2006  & XRT     &  1.30$^{+2.63}_{-1.23}$&2$^{+2.7}_{-1.1}$\\
March 2008      & XRT     &  3.0$^{+4}_{-1.54}$&4.7$^{+5.3}_{-2}$ \\
June 2008       & XRT     &  2.23$^{+2.34}_{-1.06}$& 2.8$^{+2.3}_{-1.2}$\\
Apr. 2009       & Suzaku  &  0.7\tnote{$\dagger$}&6$^{+6}_{-5}$\\
Apr. 2010       & XMM Slew & 7$^{+3}_{-3}$&NA\\
May 2013        & XRT      & 2.24$^{+1.67}_{-1.04}$&3.7$^{+1.9}_{-1.2}$\\
June 2023       & XRT      & 2.03$^{+2.76}_{-1.04}$&0.9$^{+1.7}_{-0.5}$\\
 \hline                                   
 \end{tabular}
 \begin{tablenotes}
 \item[$\dagger$] Errors not available in the original data analysis
 \end{tablenotes}
 \end{threeparttable}
 \label{x_ray_fluxes}
 \end{table}


\begin{figure}
    \centering
    \includegraphics[width=12cm]{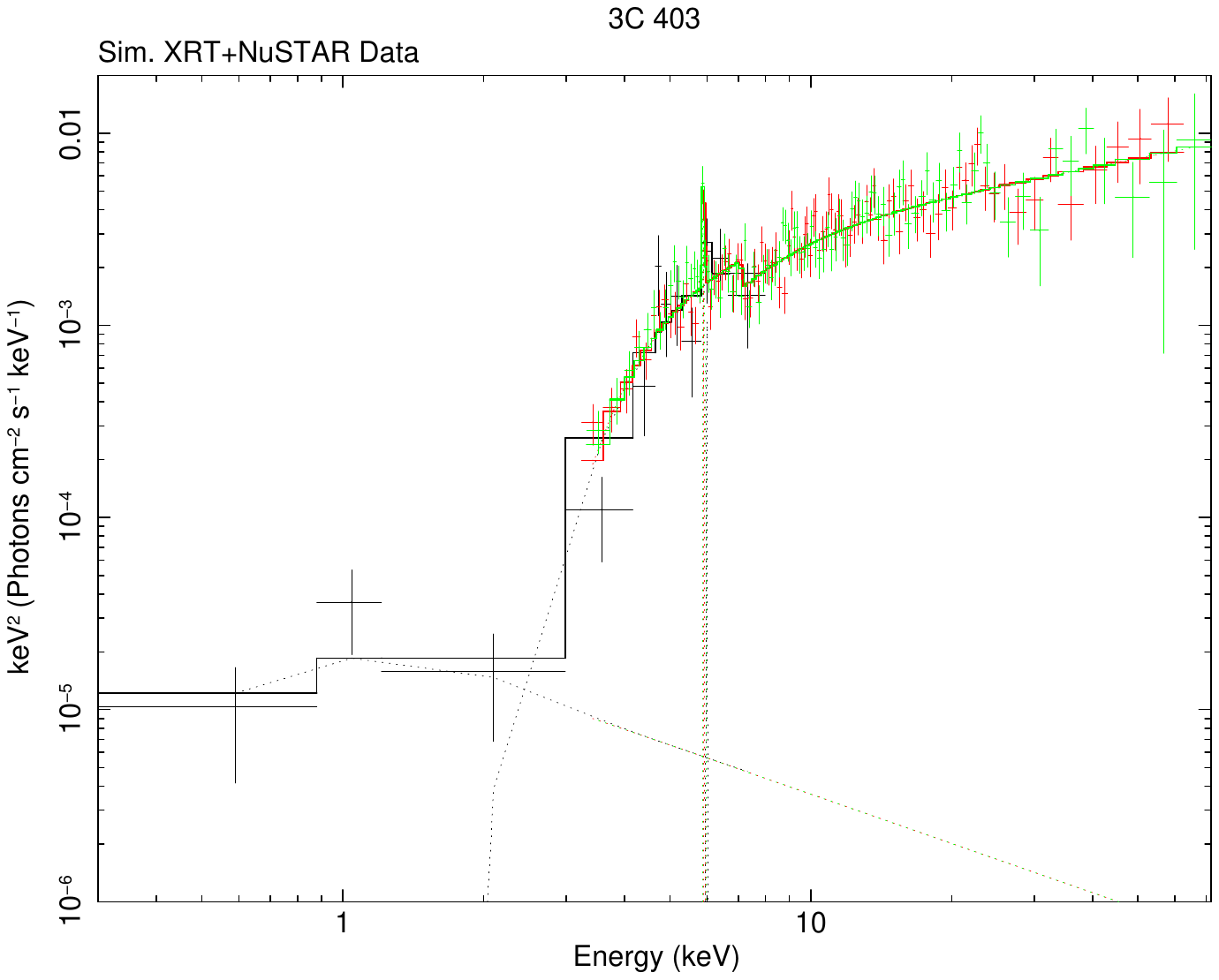}
    \caption{Simultaneous broad-band Swift XRT/NuSTAR spectral fit of 3C 403. We model the spectrum with an absorbed power law and a Gaussian component representing the Fe line, together with a second, less absorbed power law accounting for the soft X-ray excess. Galactic absorption is included in the fit.}
    \label{broad_band}
\end{figure}


\begin{figure}
    \centering
    \includegraphics[width=0.48\linewidth]{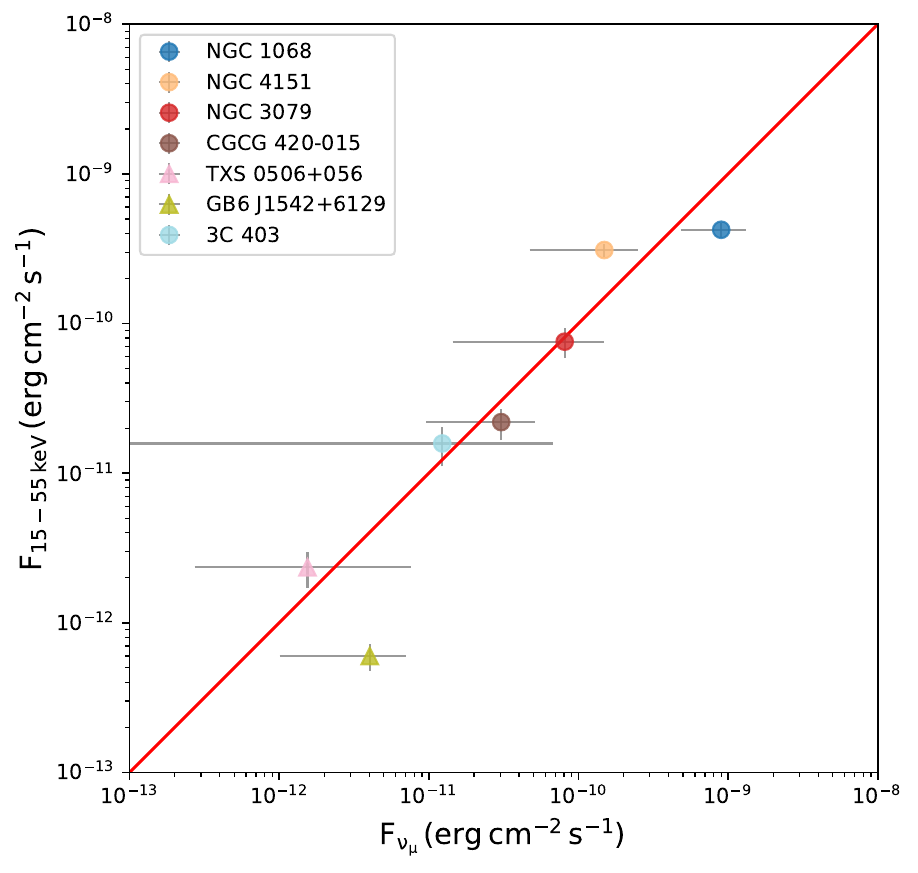}
    \includegraphics[width=0.48\linewidth]{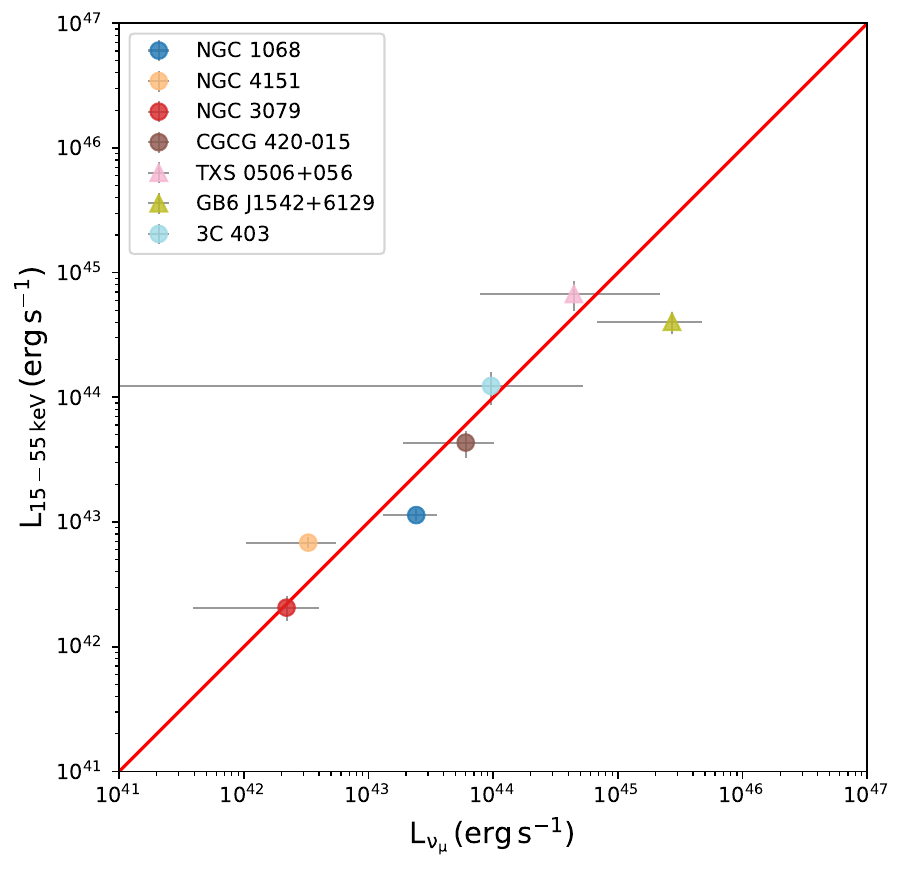}
    \caption{Adaptation of the figures from \cite{2024PhRvD.110l3014K}, with the addition of 3C\,403. The two blazars of the sample (TXS\,0506+056 and GB6\,J1542+6129) are indicated with a triangle.}
    \label{fig:Kun}
\end{figure}

\section{The proposed correlation between Hard X-rays and neutrino fluxes}
\label{kun}

Recent works have investigated the possible connection between high-energy neutrino emission and hard X-ray radiation in AGN, motivated by scenarios in which neutrinos are produced through hadronic interactions involving the same regions responsible for the X-ray emission. In particular, models based on accretion-flow coronae predict a relation between neutrino luminosity and hard X-ray luminosity, as both trace the dissipated power in the inner accretion flow \citep{PhysRevD.94.103006,2024PhRvD.110l3014K}. In this framework, the $L_{\nu}$--$L_{\rm hX}$ plane provides a natural diagnostic to compare different source classes and to assess the viability of coronal neutrino production.

Before discussing the implications of the $L_{\nu}$--$L_{\rm hX}$ plane for neutrino production, it is useful to place 3C\,403 in the broader context of AGN accretion regimes. In the optical spectroscopic classification of the 3CR sample, 3C\,403 is identified as a high-excitation galaxy (HEG) \citep{2010A&A...509A...6B}, a class generally associated with radiatively efficient accretion in radio-loud AGN \citep[e.g.][]{2012MNRAS.421.1569B}. 
Adopting a black-hole mass of $\log(M_{\bullet}/M_{\odot})\simeq 8.8$, as reported for 3C\,403 on the basis of host-galaxy scaling relations \citep{Bettoni2003}, we estimate an Eddington luminosity $L_{\rm Edd}\simeq 8\times10^{46}\,\mathrm{erg\,s^{-1}}$.
Using an average intrinsic (absorption-corrected) $2$--$10$~keV flux $F_{2-10}\sim6\times10^{-12}\,\mathrm{erg\,cm^{-2}\,s^{-1}}$ (Table~\ref{x_ray_fluxes}), we obtain an intrinsic luminosity $L_{2-10}\sim5\times10^{43}\,\mathrm{erg\,s^{-1}}$. 
Adopting a standard X-ray bolometric correction $k_{\rm bol}\sim15$--$25$ \citep{Marconi2004,Duras2020}, we estimate a bolometric luminosity $L_{\rm bol}\sim(0.8$--$1.3)\times10^{45}\,\mathrm{erg\,s^{-1}}$, corresponding to an Eddington ratio $\lambda_{\rm Edd}\sim10^{-2}$. 
The dominant uncertainties arise from the bolometric correction and the black-hole mass estimate.

This value places 3C\,403 well within the range spanned by typical Seyfert galaxies, which commonly exhibit $\lambda_{\rm Edd}\sim10^{-3}$--$10^{-1}$ \citep[e.g.][]{2009MNRAS.399..349G,2009MNRAS.392.1124V}. 
At the same time, it lies close to the commonly discussed transition between radiatively efficient and inefficient accretion regimes, often placed around $\lambda_{\rm Edd}\sim10^{-2}$, the latter being typically associated with BL\,Lac objects \citep{2009MNRAS.396L.105T}.
We note that the intermediate location of 3C\,403 in the $L_{\nu}$--$L_{\rm hX}$ plane is unlikely to be primarily driven by black-hole mass. 
Its estimated mass lies within the broad range observed in both local Seyfert galaxies ($\sim10^{6}$--$10^{8}\,M_{\odot}$; \cite{Peterson2004,Bentz2015,Ricci2017}) and radio-loud AGN and blazars ($\sim10^{8}$--$10^{9}\,M_{\odot}$; \cite{Wu2009,Shaw2012,Ghisellini2010}), in the region where the two distributions overlap. 
This suggests that black-hole mass alone is unlikely to account for its intermediate position in the diagram, and that differences in emission geometry and in the physical origin of the hard X-ray radiation likely play a more significant role.
While this does not constitute evidence for neutrino emission from 3C\,403, it provides a physically motivated context for interpreting its location in the $L_{\nu}$--$L_{\rm hX}$ plane. In that observational plane, 3C\,403 occupies a position between jet-dominated blazars and corona-dominated Seyfert galaxies, reflecting its combination of strong hard X-ray emission and a powerful but misaligned radio jet.

The comparison also highlights that the physical origin of the hard X-ray emission differs substantially across AGN classes. In Seyfert galaxies and misaligned radio galaxies, hard X-rays predominantly trace the accretion-flow corona, whereas in BL\,Lac objects they are dominated by Doppler-boosted jet emission. Despite this diversity, all sources included in the $L_{\nu}$--$L_{\rm hX}$ plane share the common property of being hard-X–bright.
This suggests a more general, observationally driven interpretation in which the presence of an intense hard X-ray photon field constitutes a key ingredient for efficient neutrino production, largely independent of whether such photons originate in the corona or in the jet. In this picture, hard X-rays primarily trace the photon target density required for hadronic interactions, rather than uniquely identifying the physical site of neutrino production.

Within this framework, the location of neutrino production is expected to be class-dependent. In Seyfert galaxies and misaligned radio galaxies, where Doppler boosting is negligible and the hard X-ray emission is dominated by the corona, neutrinos may be produced in the corona itself or at the base of the jet, where relativistic protons can efficiently interact with coronal X-ray photons. In BL\,Lac objects, instead, hard X-rays are produced within the relativistic jet and are strongly Doppler-boosted, naturally favouring jet-based neutrino production scenarios operating at the same dissipation sites responsible for the observed high-energy emission.

The lack of a clear neutrino detection from 3C\,403 implies that its position in the $L_{\nu}$--$L_{\rm hX}$ plane remains observationally unconstrained, and its interpretation as a transitional source in this diagnostic diagram should therefore be regarded as tentative. Nevertheless, its hard X-ray brightness, misaligned jet geometry, and radiatively efficient accretion properties make 3C\,403 a physically motivated candidate for probing the connection between coronal and jet environments in future multimessenger studies.

\section{Conclusions}

We presented a multi-wavelength study of 3C\,403, a well-known FR\,II radio galaxy identified as a candidate neutrino emitter in the 15-year ANTARES dataset.
Our analysis of the radio morphology from kiloparsec to parsec scales, based on VLA, e-MERLIN, and VLBA observations, reveals a symmetric two-sided jet with no evidence of strong Doppler beaming or large-scale precession.
The jet lies close to the plane of the sky and appears stable over a $\sim$Myr timescale, implying efficient coupling with the surrounding interstellar medium.
Such conditions are favourable to the formation of backflows and shocks, which may enhance hadronic ($pp$) interactions, as proposed for other X-shaped radio galaxies \citep{Cotton2020}.

At higher energies, 3C\,403 displays a persistent, heavily absorbed hard X-ray continuum dominated by accretion-related emission, with no evidence for strong variability. 
These characteristics indicate the presence of a long-lived and energetically significant hard X-ray photon field, providing favourable conditions for hadronic interactions involving accelerated protons, irrespective of the specific physical site where such interactions may occur. 
While coronal $p\gamma$ scenarios have been proposed for sources such as NGC\,1068, and jet-based models for blazars like TXS\,0506+056 \citep{Inoue2020,Fiorillo2024b,Fiorillo2025}, the case of 3C\,403 highlights the importance of considering both accretion-related and jet-related environments within a unified observational framework.

By placing 3C\,403 in the context of the hard X-ray versus neutrino flux and luminosity planes \citep{2024PhRvD.110l3014K}, we explored a complementary, observationally motivated interpretation in which the hard X-ray emission acts as a tracer of the high-energy photon fields required for neutrino production. 
In this view, intense hard X-ray emission emerges as a common property of candidate neutrino-emitting AGN, while the physical site of neutrino production—corona, jet base, or relativistic jet—may depend on source class and geometry.

Although no neutrino detection is currently available for 3C\,403, its hard X-ray brightness, radiatively efficient accretion ($\lambda_{\rm Edd}\sim10^{-2}$), and powerful but misaligned jet place it in a region of the $L_{\nu}$--$L_{\rm hX}$ plane between jet-dominated blazars and corona-dominated Seyfert galaxies. 
This makes 3C\,403 a physically motivated, yet currently unconstrained, laboratory for testing neutrino production scenarios across different AGN classes.

Future neutrino observatories with improved sensitivity will be crucial to assess whether misaligned radio galaxies such as 3C\,403 can contribute to the observed astrophysical neutrino flux and to clarify the relative roles of jets and accretion flows in powering these cosmic messengers.


\begin{acknowledgments}
The authors who are not members of the ANTARES Collaboration gratefully acknowledge the Collaboration for providing access to private information on the results of the 15-year data taking.
We thank Emma Kun for the useful discussion, and for kindly providing the data to reproduce the figure from her work. 
GB acknowledges financial support for the GRACE project, selected through the Open Space Innovation Platform (\url{https://ideas.esa.int}) as a Co-Sponsored Research Agreement and carried out under the Discovery program of and funded by the European Space Agency (agreement No.
4000142106/23/NL/MGu/my).
GB acknowledges financial support from the Bando Ricerca Fondamentale INAF 2023 for the project: \textit{\lq\lq The GRACE project: high-energy giant radio galaxies and their duty cycle\rq\rq}. 
JR and GB acknowledge financial support under the \textit{INTEGRAL} ASI/INAF No. 2019-35.HH.0 and funding from the European Union’s Horizon 2020 Programme under the AHEAD2020 project (grant agreement n. 871158).
MF acknowledges financial support from the Bando Ricerca Fondamentale INAF 2024 for the Guest Observer Grant: \textit{\lq\lq INTEGRAL view of the Galactic Plane\rq\rq}.
The National Radio Astronomy Observatory is a facility of the National Science Foundation operated under cooperative agreement by Associated Universities, Inc.  This work made use of the Swinburne University of Technology software correlator, developed as part of the Australian Major National Research Facilities Programme and operated under licence.
e-MERLIN is a National Facility operated by the University of Manchester at Jodrell Bank Observatory on behalf of STFC, part of UK Research and Innovation.
\end{acknowledgments}

\appendix

\newcommand{\dd}[1]{\mathrm{d}{#1}}

\section{Observations and data reduction}
\label{data}
In the following, we report details about multiwavelength data from the different instruments used in this work.

\subsection{ANTARES}
\label{antares}
The search for neutrino sources performed by the ANTARES collaboration is based on a maximum-likelihood method that exploits the differences between cosmic and atmospheric neutrinos in terms of energy and spatial clustering. The only free parameter in the likelihood is the number of signal events hidden in the data, for which the best estimate, $\hat{\mu}_\text{sig}$, is found through maximization. In the case of 3C\,403, the method yields $\hat{\mu}_\text{sig} = 2.47$ for a neutrino energy spectrum $\propto E^{-2}$. Number of signal events and neutrino fluxes are related through the following equation:
\begin{equation}\label{eq_Aeff}
    n_s = \sum_{\nu_e,\nu_\mu,\nu_\tau} \int A^{\nu + \bar{\nu}}_{\rm eff}(E)\times \phi_0^{\nu + \bar{\nu}} \left( \frac{E}{E_0}\right)^{-\gamma} \dd{E},
\end{equation}
where $A^{\nu + \bar{\nu}}_{\rm eff}$ is the effective area of the detector for a given neutrino flavor and with the spectral index $\gamma$ set to $2$. Thus, given that the ANTARES fit is completely driven by track-like events, which are mainly associated to $\nu_\mu+\bar{\nu}_\mu$ charged current interactions, Equation \eqref{eq_Aeff} can be used to obtain the muon neutrino flux normalization \hbox{$\phi_{\rm 1 \,TeV}^{\nu_\mu + \bar{\nu}_\mu} = 2.63 \cdot 10^{-12} \, {\rm TeV^{-1}cm^{-2}s^{-1}}$}.\\

The energy-integrated muon-neutrino flux, $F_{\nu_\mu + \bar{\nu}_\mu}$, as shown in the x-axis of Figure \ref{fig:Kun}, is straightforward to obtain as:
\begin{equation}
    F_{\nu_\mu + \bar{\nu}_\mu} = \int^{E_{\rm max}}_{E_{\rm min}} \phi_{\rm 1 \,TeV}^{\nu_\mu + \bar{\nu}_\mu} \left( \frac{E}{\rm 1\, TeV}\right)^{-2} \times E \times \dd{E},
\end{equation}
with $E_{\rm max} = 100 \, \rm TeV$ and $E_{\rm min} = 0.3 \, \rm TeV$, as in \cite{2024PhRvD.110l3014K}. After the proper unit transformation is done, an energy-integrated flux $F_{\nu_\mu + \bar{\nu}_\mu} = 2.45\cdot10^{-11}\, \rm erg\,cm^{-2}s^{-1}$ is obtained, and later divided by two to remove the contribution from antineutrinos. Since the significance of its observation is not high the ANTARES collaboration will only provide an upper limit on the neutrino flux, which is shown in Figure~\ref{fig:Kun}, as the right extreme of the horizontal error line.
The upper limit corresponds to a 90\% confidence level (CL).

\subsection{e-MERLIN}
Observations with the e-Merlin at 1.5 GHz (L-band) and 5 GHz (C-band) were carried out on November 19-20, 2022, under project CY14219 (PI Bruni). Phase referencing was applied, for a total observing time of about 12 hours for each band. Data were calibrated with the e-MERLIN pipeline \citep{2021ascl.soft09006M}, and imaged in DIFMAP. The obtained images have an RMS of 140 $\mu$Jy/beam and an angular resolution of 310$\times$138 milli-arcsec (P.A. 22 deg) at 1.5 GHz, while an RMS of 45 $\mu$Jy/beam and an angular resolution of 95$\times$41 milli-arcseconds (P.A. 21 deg) at 5 GHz.

\subsection{VLA}
We collected archival data from the NRAO VLA Archive Survey (NVAS\footnote{\href{https://www.vla.nrao.edu/astro/nvas/}{https://www.vla.nrao.edu/astro/nvas/}}). In particular, we used observations from June 10, 1994, performed in B configuration at L band (1.5 GHz). The angular resolution is 4.56$\times$4.19 arcseconds (P.A. $-$17 deg), and the RMS is 170 $\mu$Jy/beam.

\subsection{VLBA}
Observations with the Very Long Baseline Array (VLBA) were conducted on September 18, 2021, at three frequencies: 5 GHz (C-band), 8 GHz (X-band), and 15 GHz (Ku-band). The total observing time was 5.5 hours. Phase-referencing was applied, using J1951+0134 as calibrator, while 3C\,345 was used as fringe finder. Data were processed through standard \texttt{AIPS} procedures. Then, calibrated visibilities were exported, and images produced in \texttt{DIFMAP}. The angular resolution of the three different images was 3.1$\times$1.5 mas (P.A. 0.5 deg) at 5 GHz, 1.8$\times$0.9 mas (P.A. --1.9) at 8 GHz, and 1.1$\times$0.5 mas (P.A. --1.4 deg) at 15 GHz. The RMS was 48 $\mu$Jy/beam, 120 $\mu$Jy/beam, and 110 $\mu$Jy/beam, respectively.


\begin{table}[]
    \centering
    \begin{tabular}{ccccccccccc}
    \hline
    Telescope   & Date          & Frequency & FWHM              & P.A. \\ 
                & (dd/mm/yyyy)  & (GHz)     & (mas$\times$mas)  & (deg) \\
    \hline
        e-Merlin    & 19/11/2022 & 1.5  & 310$\times$138 & 22   \\
                    & 20/11/2022 & 5    & 95$\times$41   & 21   \\
        VLBA        & 18/09/2021 & 4.8  & 3.1$\times$1.5 & 0.5  \\
                    & 18/09/2021 & 8.4  & 1.8$\times$0.9 & --1.9\\
                    & 18/09/2021 & 15   & 1.1$\times$0.5 & --1.4\\
    \hline
    \end{tabular}
    \caption{Radio observations log. The Full Width Half Maximum (FWHM) of the PSF main lobe, generally dubbed \textit{beam}, and the position angle (P.A.) of its major axis are given.}
    \label{tab:radio}
\end{table}

\subsection{\textit{INTEGRAL}/IBIS}

The analyzed \textit{INTEGRAL}/IBIS data consist of all observations started on 2003-03-09  and ended 2024-09-26  within the 12° of high-energy detectors, for a total of 2.9 Ms. \textit{INTEGRAL}/IBIS \citep{2003A&A...411L.131U} data are processed using the Off-line Scientific Analysis (OSA v11.2) software released by the \textit{INTEGRAL} Scientific Data Centre \citep{2003A&A...411L..53C}.

\subsection{\textit{NuSTAR}}

\textit{NuSTAR} data (from both focal plane detectors, FPMA and FPMB)
were reduced using the \texttt{nustardas$\_$04May21$\_$v2.1.1} and \texttt{CALDB} version
20220118. For 3C\,403, In this work, we reduced observation 60061293002, taken May 25, 2013,
with a cleaned exposure of $\sim$20 ksec; spectral extraction and the subsequent production of
response and ancillary files were performed using the
\texttt{nuproducts} task with an extraction radius of $\sim$40$^{\prime\prime}$; 
to maximise the signal-to-noise ratio; background spectra were extracted from 
circular regions of 40$^{\prime\prime}$ radius 
in source-free areas of the detectors. 


\begin{table}

\centering
\caption{XRT-NuSTAR Broad-band spectral analysis}
\begin{tabular}{lr}
\hline
\multicolumn{2}{c}{\texttt{const*phabs[po+phabs*(po+zga)]}}\\
\hline
 N$_{\rm H}$& (22.24$^{+3.55}_{-3.20}$)$\times$10$^{22}$cm$^{-2}$\\
  $\Gamma_{\rm cont}$&1.52$\pm$0.09\\
  $\Gamma_{\rm soft}$&2.85$^{+1.96}_{-1.45}$\\
 E$_{\rm Fe}$ (k$\alpha$)& 6.25$\pm$0.09 keV\\
 EW& 201$^{+92}_{-93}$ eV\\
 F$_{2-10}$& 2.57$\times$10$^{-12}$ erg cm$^{-2}$ s$^{-1}$ \\
 F$_{20-100}$& 1.85$\times$10$^{-11}$ erg cm$^{-2}$ s$^{-1}$  \\
$\chi^{2}$ (d.o.f.)& 191.35 (189)\\  
\hline

\end{tabular}

\label{best_fit}
\end{table}


\bibliography{./3C403.bib}{}

\providecommand{\href}[2]{#2}\begingroup\raggedright\begin{thebibliography}{10}

\bibitem{PhysRevLett.111.021103}
{\scshape IceCube Collaboration} collaboration, \emph{First observation of pev-energy neutrinos with icecube}, \href{https://doi.org/10.1103/PhysRevLett.111.021103}{\emph{Phys. Rev. Lett.} {\bfseries 111} (2013) 021103}.

\bibitem{doi:10.1126/science.aat2890}
I.~Collaboration, M.~Aartsen, M.~Ackermann, J.~Adams, J.A.~Aguilar, M.~Ahlers et~al., \emph{Neutrino emission from the direction of the blazar txs 0506+056 prior to the icecube-170922a alert}, \href{https://doi.org/10.1126/science.aat2890}{\emph{Science} {\bfseries 361} (2018) 147} [\href{https://arxiv.org/abs/https://www.science.org/doi/pdf/10.1126/science.aat2890}{{\ttfamily https://www.science.org/doi/pdf/10.1126/science.aat2890}}].

\bibitem{2019NatAs...3...88G}
S.~{Gao}, A.~{Fedynitch}, W.~{Winter} and M.~{Pohl}, \emph{{Modelling the coincident observation of a high-energy neutrino and a bright blazar flare}}, \href{https://doi.org/10.1038/s41550-018-0610-1}{\emph{Nature Astronomy} {\bfseries 3} (2019) 88} [\href{https://arxiv.org/abs/1807.04275}{{\ttfamily 1807.04275}}].

\bibitem{2019ApJ...880..103G}
S.~{Garrappa}, S.~{Buson}, A.~{Franckowiak}, {Fermi-LAT Collaboration}, B.J.~{Shappee}, J.F.~{Beacom} et~al., \emph{{Investigation of Two Fermi-LAT Gamma-Ray Blazars Coincident with High-energy Neutrinos Detected by IceCube}}, \href{https://doi.org/10.3847/1538-4357/ab2ada}{\emph{\apj} {\bfseries 880} (2019) 103} [\href{https://arxiv.org/abs/1901.10806}{{\ttfamily 1901.10806}}].

\bibitem{1989A&A...221..211M}
K.~{Mannheim} and P.L.~{Biermann}, \emph{{Photomeson production in active galactic nuclei.}}, {\emph{\aap} {\bfseries 221} (1989) 211}.

\bibitem{1990ApJ...362...38B}
M.C.~{Begelman}, B.~{Rudak} and M.~{Sikora}, \emph{{Consequences of Relativistic Proton Injection in Active Galactic Nuclei}}, \href{https://doi.org/10.1086/169241}{\emph{\apj} {\bfseries 362} (1990) 38}.

\bibitem{PhysRevLett.116.071101}
K.~Murase, D.~Guetta and M.~Ahlers, \emph{Hidden cosmic-ray accelerators as an origin of tev-pev cosmic neutrinos}, \href{https://doi.org/10.1103/PhysRevLett.116.071101}{\emph{Phys. Rev. Lett.} {\bfseries 116} (2016) 071101}.

\bibitem{PhysRevLett.125.011101}
K.~Murase, S.S.~Kimura and P.~M\'esz\'aros, \emph{Hidden cores of active galactic nuclei as the origin of medium-energy neutrinos: Critical tests with the mev gamma-ray connection}, \href{https://doi.org/10.1103/PhysRevLett.125.011101}{\emph{Phys. Rev. Lett.} {\bfseries 125} (2020) 011101}.

\bibitem{PhysRevLett.132.101002}
A.~Neronov, D.~Savchenko and D.V.~Semikoz, \emph{Neutrino signal from a population of seyfert galaxies}, \href{https://doi.org/10.1103/PhysRevLett.132.101002}{\emph{Phys. Rev. Lett.} {\bfseries 132} (2024) 101002}.

\bibitem{Murase_2022}
K.~Murase, \emph{Hidden hearts of neutrino active galaxies}, \href{https://doi.org/10.3847/2041-8213/aca53c}{\emph{The Astrophysical Journal Letters} {\bfseries 941} (2022) L17}.

\bibitem{1997A&A...325L..13M}
G.~{Matt}, M.~{Guainazzi}, F.~{Frontera}, L.~{Bassani}, W.N.~{Brandt}, A.C.~{Fabian} et~al., \emph{{Hard X-ray detection of NGC 1068 with BeppoSAX.}}, \href{https://doi.org/10.48550/arXiv.astro-ph/9707065}{\emph{\aap} {\bfseries 325} (1997) L13} [\href{https://arxiv.org/abs/astro-ph/9707065}{{\ttfamily astro-ph/9707065}}].

\bibitem{2015ApJ...812..116B}
F.E.~{Bauer}, P.~{Ar{\'e}valo}, D.J.~{Walton}, M.J.~{Koss}, S.~{Puccetti}, P.~{Gandhi} et~al., \emph{{NuSTAR Spectroscopy of Multi-component X-Ray Reflection from NGC 1068}}, \href{https://doi.org/10.1088/0004-637X/812/2/116}{\emph{\apj} {\bfseries 812} (2015) 116} [\href{https://arxiv.org/abs/1411.0670}{{\ttfamily 1411.0670}}].

\bibitem{2020ApJ...891L..33I}
Y.~{Inoue}, D.~{Khangulyan} and A.~{Doi}, \emph{{On the Origin of High-energy Neutrinos from NGC 1068: The Role of Nonthermal Coronal Activity}}, \href{https://doi.org/10.3847/2041-8213/ab7661}{\emph{\apjl} {\bfseries 891} (2020) L33} [\href{https://arxiv.org/abs/1909.02239}{{\ttfamily 1909.02239}}].

\bibitem{2021ApJ...922...45K}
A.~{Kheirandish}, K.~{Murase} and S.S.~{Kimura}, \emph{{High-energy Neutrinos from Magnetized Coronae of Active Galactic Nuclei and Prospects for Identification of Seyfert Galaxies and Quasars in Neutrino Telescopes}}, \href{https://doi.org/10.3847/1538-4357/ac1c77}{\emph{\apj} {\bfseries 922} (2021) 45} [\href{https://arxiv.org/abs/2102.04475}{{\ttfamily 2102.04475}}].

\bibitem{2024ApJ...972...44D}
A.~{Das}, B.T.~{Zhang} and K.~{Murase}, \emph{{Revealing the Production Mechanism of High-energy Neutrinos from NGC 1068}}, \href{https://doi.org/10.3847/1538-4357/ad5a04}{\emph{\apj} {\bfseries 972} (2024) 44} [\href{https://arxiv.org/abs/2405.09332}{{\ttfamily 2405.09332}}].

\bibitem{2024PhRvD.109j1306M}
R.~{Mbarek}, A.~{Philippov}, A.~{Chernoglazov}, A.~{Levinson} and R.~{Mushotzky}, \emph{{Interplay between accelerated protons, x rays and neutrinos in the corona of NGC 1068: Constraints from kinetic plasma simulations}}, \href{https://doi.org/10.1103/PhysRevD.109.L101306}{\emph{\prd} {\bfseries 109} (2024) L101306} [\href{https://arxiv.org/abs/2310.15222}{{\ttfamily 2310.15222}}].

\bibitem{2025arXiv250406336F}
D.F.G.~{Fiorillo}, L.~{Comisso}, E.~{Peretti}, M.~{Petropoulou} and L.~{Sironi}, \emph{{The contribution of turbulent AGN coronae to the diffuse neutrino flux}}, \href{https://doi.org/10.48550/arXiv.2504.06336}{\emph{arXiv e-prints} (2025) arXiv:2504.06336} [\href{https://arxiv.org/abs/2504.06336}{{\ttfamily 2504.06336}}].

\bibitem{2024ApJ...961L..14F}
D.F.G.~{Fiorillo}, M.~{Petropoulou}, L.~{Comisso}, E.~{Peretti} and L.~{Sironi}, \emph{{TeV Neutrinos and Hard X-Rays from Relativistic Reconnection in the Corona of NGC 1068}}, \href{https://doi.org/10.3847/2041-8213/ad192b}{\emph{\apjl} {\bfseries 961} (2024) L14} [\href{https://arxiv.org/abs/2310.18254}{{\ttfamily 2310.18254}}].

\bibitem{2024ApJ...974...75F}
D.F.G.~{Fiorillo}, L.~{Comisso}, E.~{Peretti}, M.~{Petropoulou} and L.~{Sironi}, \emph{{A Magnetized Strongly Turbulent Corona as the Source of Neutrinos from NGC 1068}}, \href{https://doi.org/10.3847/1538-4357/ad7021}{\emph{\apj} {\bfseries 974} (2024) 75} [\href{https://arxiv.org/abs/2407.01678}{{\ttfamily 2407.01678}}].

\bibitem{2025arXiv250201738F}
D.F.G.~{Fiorillo}, F.~{Testagrossa}, M.~{Petropoulou} and W.~{Winter}, \emph{{Can the neutrinos from TXS 0506+056 have a coronal origin?}}, \href{https://doi.org/10.48550/arXiv.2502.01738}{\emph{arXiv e-prints} (2025) arXiv:2502.01738} [\href{https://arxiv.org/abs/2502.01738}{{\ttfamily 2502.01738}}].

\bibitem{PhysRevD.94.103006}
K.~Murase and E.~Waxman, \emph{Constraining high-energy cosmic neutrino sources: Implications and prospects}, \href{https://doi.org/10.1103/PhysRevD.94.103006}{\emph{Phys. Rev. D} {\bfseries 94} (2016) 103006}.

\bibitem{2024PhRvD.110l3014K}
E.~{Kun}, I.~{Bartos}, J.B.~{Tjus}, P.L.~{Biermann}, A.~{Franckowiak}, F.~{Halzen} et~al., \emph{{Possible correlation between unabsorbed hard x rays and neutrinos in radio-loud and radio-quiet active galactic nuclei}}, \href{https://doi.org/10.1103/PhysRevD.110.123014}{\emph{\prd} {\bfseries 110} (2024) 123014} [\href{https://arxiv.org/abs/2404.06867}{{\ttfamily 2404.06867}}].

\bibitem{2019ATel12967....1T}
I.~{Taboada} and R.~{Stein}, \emph{{IceCube-190730A an astrophysical neutrino candidate in spatial coincidence with FSRQ PKS 1502+106}}, {\emph{The Astronomer's Telegram} {\bfseries 12967} (2019) 1}.

\bibitem{2016NatPh..12..807K}
M.~{Kadler}, F.~{Krau{\ss}}, K.~{Mannheim}, R.~{Ojha}, C.~{M{\"u}ller}, R.~{Schulz} et~al., \emph{{Coincidence of a high-fluence blazar outburst with a PeV-energy neutrino event}}, \href{https://doi.org/10.1038/nphys3715}{\emph{Nature Physics} {\bfseries 12} (2016) 807} [\href{https://arxiv.org/abs/1602.02012}{{\ttfamily 1602.02012}}].

\bibitem{2024arXiv240607601A}
R.~{Abbasi}, M.~{Ackermann}, J.~{Adams}, S.K.~{Agarwalla}, J.A.~{Aguilar}, M.~{Ahlers} et~al., \emph{{IceCube Search for Neutrino Emission from X-ray Bright Seyfert Galaxies}}, \href{https://doi.org/10.48550/arXiv.2406.07601}{\emph{arXiv e-prints} (2024) arXiv:2406.07601} [\href{https://arxiv.org/abs/2406.07601}{{\ttfamily 2406.07601}}].

\bibitem{Ambrosone2024}
A.~Ambrosone, \emph{Berezinsky hidden sources: an emergent tension in the high-energy neutrino sky?}, \href{https://doi.org/10.1088/1475-7516/2024/09/075}{\emph{Journal of Cosmology and Astroparticle Physics} {\bfseries 2024} (2024) 075}.

\bibitem{2016MNRAS.461.3165B}
L.~{Bassani}, T.~{Venturi}, M.~{Molina}, A.~{Malizia}, D.~{Dallacasa}, F.~{Panessa} et~al., \emph{{Soft {\ensuremath{\gamma}}-ray selected radio galaxies: favouring giant size discovery}}, \href{https://doi.org/10.1093/mnras/stw1468}{\emph{\mnras} {\bfseries 461} (2016) 3165} [\href{https://arxiv.org/abs/1606.05456}{{\ttfamily 1606.05456}}].

\bibitem{2021ApJ...911...48A}
A.~{Albert}, M.~{Andr{\'e}}, M.~{Anghinolfi}, G.~{Anton}, M.~{Ardid}, J.J.~{Aubert} et~al., \emph{{ANTARES Search for Point Sources of Neutrinos Using Astrophysical Catalogs: A Likelihood Analysis}}, \href{https://doi.org/10.3847/1538-4357/abe53c}{\emph{\apj} {\bfseries 911} (2021) 48} [\href{https://arxiv.org/abs/2012.15082}{{\ttfamily 2012.15082}}].

\bibitem{AlvesPhD_2025}
S.~Alves~Garre, \emph{Search for the Origin of Cosmic Rays with the ANTARES Neutrino Telescope}, Ph.D. thesis, Univesidad de Valencia, 2025.

\bibitem{2003A&A...407L..33T}
A.~{Tarchi}, C.~{Henkel}, M.~{Chiaberge} and K.M.~{Menten}, \emph{{Discovery of a luminous water megamaser in the FRII radiogalaxy 3C 403}}, \href{https://doi.org/10.1051/0004-6361:20031062}{\emph{\aap} {\bfseries 407} (2003) L33} [\href{https://arxiv.org/abs/astro-ph/0307068}{{\ttfamily astro-ph/0307068}}].

\bibitem{2007A&A...475..497T}
A.~{Tarchi}, A.~{Brunthaler}, C.~{Henkel}, K.M.~{Menten}, J.~{Braatz} and A.~{Wei{\ss}}, \emph{{The innermost region of the water megamaser radio galaxy 3C 403}}, \href{https://doi.org/10.1051/0004-6361:20078317}{\emph{\aap} {\bfseries 475} (2007) 497} [\href{https://arxiv.org/abs/0709.3417}{{\ttfamily 0709.3417}}].

\bibitem{2020A&A...641A.162P}
F.~{Panessa}, P.~{Castangia}, A.~{Malizia}, L.~{Bassani}, A.~{Tarchi}, A.~{Bazzano} et~al., \emph{{Water megamaser emission in hard X-ray selected AGN}}, \href{https://doi.org/10.1051/0004-6361/201937407}{\emph{\aap} {\bfseries 641} (2020) A162} [\href{https://arxiv.org/abs/2006.08280}{{\ttfamily 2006.08280}}].

\bibitem{2023AAS...24125407L}
A.~{Lien}, H.~{Krimm}, C.~{Markwardt}, N.~{Collins}, S.~{Barthelmy}, K.~{Oh} et~al., \emph{{The BAT 157 month survey catalog and beyond}},  in \emph{American Astronomical Society Meeting Abstracts}, vol.~241 of \emph{American Astronomical Society Meeting Abstracts}, p.~254.07, Jan., 2023.

\bibitem{Bird_2016}
A.J.~Bird, A.~Bazzano, A.~Malizia, M.~Fiocchi, V.~Sguera, L.~Bassani et~al., \emph{The ibis soft gamma-ray sky after 1000 integral orbits*}, \href{https://doi.org/10.3847/0067-0049/223/1/15}{\emph{The Astrophysical Journal Supplement Series} {\bfseries 223} (2016) 15}.

\bibitem{Abdollahi_2022}
S.~Abdollahi, F.~Acero, L.~Baldini, J.~Ballet, D.~Bastieri, R.~Bellazzini et~al., \emph{Incremental fermi large area telescope fourth source catalog}, \href{https://doi.org/10.3847/1538-4365/ac6751}{\emph{The Astrophysical Journal Supplement Series} {\bfseries 260} (2022) 53}.

\bibitem{2007AJ....133.2097C}
C.C.~{Cheung}, \emph{{FIRST ``Winged'' and X-Shaped Radio Source Candidates}}, \href{https://doi.org/10.1086/513095}{\emph{\aj} {\bfseries 133} (2007) 2097} [\href{https://arxiv.org/abs/astro-ph/0701278}{{\ttfamily astro-ph/0701278}}].

\bibitem{2018ApJ...852...48S}
L.~{Saripalli} and D.H.~{Roberts}, \emph{{What Are {\textquotedblleft}X-shaped{\textquotedblright} Radio Sources Telling Us? II. Properties of a Sample of 87}}, \href{https://doi.org/10.3847/1538-4357/aa9c4b}{\emph{\apj} {\bfseries 852} (2018) 48} [\href{https://arxiv.org/abs/1710.01652}{{\ttfamily 1710.01652}}].

\bibitem{1980Natur.287..307B}
M.C.~{Begelman}, R.D.~{Blandford} and M.J.~{Rees}, \emph{{Massive black hole binaries in active galactic nuclei}}, \href{https://doi.org/10.1038/287307a0}{\emph{\nat} {\bfseries 287} (1980) 307}.

\bibitem{2020MNRAS.495.1271C}
W.D.~{Cotton}, K.~{Thorat}, J.J.~{Condon}, B.S.~{Frank}, G.I.G.~{J{\'o}zsa}, S.V.~{White} et~al., \emph{{Hydrodynamical backflow in X-shaped radio galaxy PKS 2014-55}}, \href{https://doi.org/10.1093/mnras/staa1240}{\emph{\mnras} {\bfseries 495} (2020) 1271} [\href{https://arxiv.org/abs/2005.02723}{{\ttfamily 2005.02723}}].

\bibitem{1985A&AS...59..511P}
P.~{Parma}, R.D.~{Ekers} and R.~{Fanti}, \emph{{High resolution radio observations of low luminosity radio galaxies.}}, {\emph{\aaps} {\bfseries 59} (1985) 511}.

\bibitem{1994A&AS..103..157M}
K.H.~{Mack}, L.~{Gregorini}, P.~{Parma} and U.~{Klein}, \emph{{High-frequency radio continuum observations of radio galaxies with low and intermediate luminosity. II. Sources with sizes 4' to 5'}}, {\emph{\aaps} {\bfseries 103} (1994) 157}.

\bibitem{2002MNRAS.330..609D}
J.~{Dennett-Thorpe}, P.A.G.~{Scheuer}, R.A.~{Laing}, A.H.~{Bridle}, G.G.~{Pooley} and W.~{Reich}, \emph{{Jet reorientation in active galactic nuclei: two winged radio galaxies}}, \href{https://doi.org/10.1046/j.1365-8711.2002.05106.x}{\emph{\mnras} {\bfseries 330} (2002) 609} [\href{https://arxiv.org/abs/astro-ph/0110339}{{\ttfamily astro-ph/0110339}}].

\bibitem{2005ApJ...622..149K}
R.P.~{Kraft}, M.J.~{Hardcastle}, D.M.~{Worrall} and S.S.~{Murray}, \emph{{A Chandra Study of the Multicomponent X-Ray Emission from the X-shaped Radio Galaxy 3C 403}}, \href{https://doi.org/10.1086/427822}{\emph{\apj} {\bfseries 622} (2005) 149} [\href{https://arxiv.org/abs/astro-ph/0501031}{{\ttfamily astro-ph/0501031}}].

\bibitem{10.1093/mnras/stz1910}
M.J.~Hardcastle, J.H.~Croston, T.W.~Shimwell, C.~Tasse, G.~Gürkan, R.~Morganti et~al., \emph{Ngc 326: X-shaped no more}, \href{https://doi.org/10.1093/mnras/stz1910}{\emph{Monthly Notices of the Royal Astronomical Society} {\bfseries 488} (2019) 3416} [\href{https://arxiv.org/abs/https://academic.oup.com/mnras/article-pdf/488/3/3416/29025886/stz1910.pdf}{{\ttfamily https://academic.oup.com/mnras/article-pdf/488/3/3416/29025886/stz1910.pdf}}].

\bibitem{1998ApJ...493..632G}
G.~{Giovannini}, W.D.~{Cotton}, L.~{Feretti}, L.~{Lara} and T.~{Venturi}, \emph{{VLBI Observations of a Complete Sample of Radio Galaxies. VIII. Proper Motion in 3C 338}}, \href{https://doi.org/10.1086/305159}{\emph{\apj} {\bfseries 493} (1998) 632} [\href{https://arxiv.org/abs/astro-ph/9709287}{{\ttfamily astro-ph/9709287}}].

\bibitem{Inoue2020}
Y.~Inoue, D.~Khangulyan and A.~Doi, \emph{On the origin of high-energy neutrinos from ngc 1068: The role of nonthermal coronal activity}, \href{https://doi.org/10.3847/2041-8213/ab7661}{\emph{Astrophysical Journal Letters} {\bfseries 891} (2020) L33} [\href{https://arxiv.org/abs/1909.02239}{{\ttfamily 1909.02239}}].

\bibitem{Fiorillo2024b}
D.F.G.~Fiorillo, L.~Comisso, E.~Peretti, M.~Petropoulou and L.~Sironi, \emph{A magnetized strongly turbulent corona as the source of neutrinos from ngc 1068}, \href{https://doi.org/10.3847/1538-4357/ad7021}{\emph{Astrophysical Journal} {\bfseries 974} (2024) 75} [\href{https://arxiv.org/abs/2407.01678}{{\ttfamily 2407.01678}}].

\bibitem{10.1093/pasj/48.6.801}
S.~Miyazaki, T.~Takahashi, S.~Gunji, M.~Hirayama, T.~Kamae, Y.~Sekimoto et~al., \emph{X-ray/soft gamma-ray observation of centaurus a and its implication on the emission mechanism}, \href{https://doi.org/10.1093/pasj/48.6.801}{\emph{Publications of the Astronomical Society of Japan} {\bfseries 48} (1996) 801} [\href{https://arxiv.org/abs/https://academic.oup.com/pasj/article-pdf/48/6/801/9713563/pasj48-0801.pdf}{{\ttfamily https://academic.oup.com/pasj/article-pdf/48/6/801/9713563/pasj48-0801.pdf}}].

\bibitem{2007MNRAS.382..937M}
M.~{Molina}, M.~{Giroletti}, A.~{Malizia}, R.~{Landi}, L.~{Bassani}, A.J.~{Bird} et~al., \emph{{Broad-band X-ray spectrum of the newly discovered broad-line radio galaxy IGR J21247+5058}}, \href{https://doi.org/10.1111/j.1365-2966.2007.12464.x}{\emph{\mnras} {\bfseries 382} (2007) 937} [\href{https://arxiv.org/abs/0709.1895}{{\ttfamily 0709.1895}}].

\bibitem{2013MNRAS.428.2901W}
D.J.~{Walton}, E.~{Nardini}, A.C.~{Fabian}, L.C.~{Gallo} and R.C.~{Reis}, \emph{{Suzaku observations of `bare' active galactic nuclei}}, \href{https://doi.org/10.1093/mnras/sts227}{\emph{\mnras} {\bfseries 428} (2013) 2901} [\href{https://arxiv.org/abs/1210.4593}{{\ttfamily 1210.4593}}].

\bibitem{2020ApJ...901..111K}
J.~{Kang}, J.~{Wang} and W.~{Kang}, \emph{{NuSTAR Hard X-Ray Spectra of Radio Galaxies}}, \href{https://doi.org/10.3847/1538-4357/abadf5}{\emph{\apj} {\bfseries 901} (2020) 111} [\href{https://arxiv.org/abs/2008.03293}{{\ttfamily 2008.03293}}].

\bibitem{2019MNRAS.484.2735M}
M.~{Molina}, A.~{Malizia}, L.~{Bassani}, F.~{Ursini}, A.~{Bazzano} and P.~{Ubertini}, \emph{{Swift/XRT-NuSTAR spectra of type 1 AGN: confirming INTEGRAL results on the high-energy cut-off}}, \href{https://doi.org/10.1093/mnras/stz156}{\emph{\mnras} {\bfseries 484} (2019) 2735} [\href{https://arxiv.org/abs/1901.04188}{{\ttfamily 1901.04188}}].

\bibitem{2011ApJ...738...70T}
F.~{Tazaki}, Y.~{Ueda}, Y.~{Terashima} and R.F.~{Mushotzky}, \emph{{Suzaku View of the Swift/BAT Active Galactic Nuclei. IV. Nature of Two Narrow-line Radio Galaxies (3C 403 and IC 5063)}}, \href{https://doi.org/10.1088/0004-637X/738/1/70}{\emph{\apj} {\bfseries 738} (2011) 70} [\href{https://arxiv.org/abs/1106.2942}{{\ttfamily 1106.2942}}].

\bibitem{2010A&A...509A...6B}
S.~Buttiglione, A.~Capetti, A.~Celotti, D.J.~Axon, M.~Chiaberge, F.D.~Macchetto et~al., \emph{An optical spectroscopic survey of the 3cr sample of radio galaxies with $z<0.3$. i. presentation of the data, and spectral classification}, \href{https://doi.org/10.1051/0004-6361/200913334}{\emph{Astronomy \& Astrophysics} {\bfseries 509} (2010) A6}.

\bibitem{2012MNRAS.421.1569B}
P.N.~Best and T.M.~Heckman, \emph{On the fundamental dichotomy in the local radio-agn population: accretion, evolution and host galaxy properties}, \href{https://doi.org/10.1111/j.1365-2966.2012.20414.x}{\emph{Monthly Notices of the Royal Astronomical Society} {\bfseries 421} (2012) 1569}.

\bibitem{Bettoni2003}
D.~Bettoni, R.~Falomo, G.~Fasano and F.~Govoni, \emph{The black hole mass–bulge luminosity relation for radio galaxies}, \href{https://doi.org/10.1051/0004-6361:20021770}{\emph{Astronomy \& Astrophysics} {\bfseries 399} (2003) 869}.

\bibitem{Marconi2004}
A.~Marconi, G.~Risaliti, R.~Gilli, L.K.~Hunt, R.~Maiolino and M.~Salvati, \emph{Local supermassive black holes, relics of active galactic nuclei and the x-ray background}, \href{https://doi.org/10.1111/j.1365-2966.2004.07765.x}{\emph{Monthly Notices of the Royal Astronomical Society} {\bfseries 351} (2004) 169}.

\bibitem{Duras2020}
F.~Duras, A.~Bongiorno, F.~Ricci and et~al., \emph{The bolometric output of luminous agn}, \href{https://doi.org/10.1051/0004-6361/201936817}{\emph{Astronomy \& Astrophysics} {\bfseries 636} (2020) A73}.

\bibitem{2009MNRAS.399..349G}
M.~Gu and X.~Cao, \emph{The anticorrelation between the hard x-ray photon index and the eddington ratio in low-luminosity active galactic nuclei}, \href{https://doi.org/10.1111/j.1365-2966.2009.15277.x}{\emph{Monthly Notices of the Royal Astronomical Society} {\bfseries 399} (2009) 349}.

\bibitem{2009MNRAS.392.1124V}
R.V.~Vasudevan and A.C.~Fabian, \emph{Simultaneous x-ray/optical/uv snapshots of active galactic nuclei from xmm--newton: spectral energy distributions for the reverberation mapped sample}, \href{https://doi.org/10.1111/j.1365-2966.2008.14139.x}{\emph{Monthly Notices of the Royal Astronomical Society} {\bfseries 392} (2009) 1124}.

\bibitem{2009MNRAS.396L.105T}
F.~Tavecchio and G.~Ghisellini, \emph{Fermi blazars' divide}, \href{https://doi.org/10.1111/j.1745-3933.2009.00673.x}{\emph{Monthly Notices of the Royal Astronomical Society: Letters} {\bfseries 396} (2009) L105}.

\bibitem{Peterson2004}
B.M.~Peterson, L.~Ferrarese, K.M.~Gilbert, S.~Kaspi, M.A.~Malkan, D.~Maoz et~al., \emph{Central masses and broad-line region sizes of active galactic nuclei. ii. a homogeneous analysis of a large reverberation-mapping database}, \href{https://doi.org/10.1086/423269}{\emph{The Astrophysical Journal} {\bfseries 613} (2004) 682}.

\bibitem{Bentz2015}
M.C.~Bentz and S.~Katz, \emph{The agn black hole mass database}, \href{https://doi.org/10.1086/679601}{\emph{Publications of the Astronomical Society of the Pacific} {\bfseries 127} (2015) 67}.

\bibitem{Ricci2017}
C.~Ricci, B.~Trakhtenbrot, M.J.~Koss, Y.~Ueda, A.~Del~Moro, K.~Ichikawa et~al., \emph{Bat agn spectroscopic survey – v. x-ray properties of the swift/bat 70-month agn catalogue}, \href{https://doi.org/10.1093/mnras/stx495}{\emph{Monthly Notices of the Royal Astronomical Society} {\bfseries 468} (2017) 1273}.

\bibitem{Wu2009}
Q.~Wu and X.~Cao, \emph{The central black hole masses and eddington ratios of radio-loud active galactic nuclei}, \href{https://doi.org/10.1111/j.1365-2966.2009.15281.x}{\emph{Monthly Notices of the Royal Astronomical Society} {\bfseries 399} (2009) 349}.

\bibitem{Shaw2012}
M.S.~Shaw, R.W.~Romani, G.~Cotter, S.E.~Healey, P.F.~Michelson, A.C.S.~Readhead et~al., \emph{Spectroscopy of the largest ever gamma-ray-selected bl lac sample}, \href{https://doi.org/10.1088/0004-637X/748/1/49}{\emph{The Astrophysical Journal} {\bfseries 748} (2012) 49}.

\bibitem{Ghisellini2010}
G.~Ghisellini, F.~Tavecchio, L.~Foschini, G.~Ghirlanda, L.~Maraschi and A.~Celotti, \emph{General physical properties of bright fermi blazars}, \href{https://doi.org/10.1111/j.1365-2966.2009.15898.x}{\emph{Monthly Notices of the Royal Astronomical Society} {\bfseries 402} (2010) 497}.

\bibitem{Cotton2020}
W.D.~Cotton, K.~Thorat, J.J.~Condon, B.S.~Frank, G.I.G.~Józsa, S.V.~White et~al., \emph{Hydrodynamical backflow in x-shaped radio galaxy pks 2014-55}, \href{https://doi.org/10.1093/mnras/staa1240}{\emph{Monthly Notices of the Royal Astronomical Society} {\bfseries 495} (2020) 1271} [\href{https://arxiv.org/abs/2005.02723}{{\ttfamily 2005.02723}}].

\bibitem{Fiorillo2025}
D.F.G.~Fiorillo, F.~Testagrossa, M.~Petropoulou and W.~Winter, \emph{Can the neutrinos from txs 0506+056 have a coronal origin?}, {\emph{arXiv e-prints} (2025) } [\href{https://arxiv.org/abs/2502.01738}{{\ttfamily 2502.01738}}].

\bibitem{2021ascl.soft09006M}
J.~{Moldon}, ``{eMCP: e-MERLIN CASA pipeline}.'' Astrophysics Source Code Library, record ascl:2109.006, Sept., 2021.

\bibitem{2003A&A...411L.131U}
P.~{Ubertini}, F.~{Lebrun}, G.~{Di Cocco}, A.~{Bazzano}, A.J.~{Bird}, K.~{Broenstad} et~al., \emph{{IBIS: The Imager on-board INTEGRAL}}, \href{https://doi.org/10.1051/0004-6361:20031224}{\emph{\aap} {\bfseries 411} (2003) L131}.

\bibitem{2003A&A...411L..53C}
T.J.L.~{Courvoisier}, R.~{Walter}, V.~{Beckmann}, A.J.~{Dean}, P.~{Dubath}, R.~{Hudec} et~al., \emph{{The INTEGRAL Science Data Centre (ISDC)}}, \href{https://doi.org/10.1051/0004-6361:20031172}{\emph{\aap} {\bfseries 411} (2003) L53} [\href{https://arxiv.org/abs/astro-ph/0308047}{{\ttfamily astro-ph/0308047}}].

\end{thebibliography}\endgroup
\bibliographystyle{./JHEP}
\end{document}